\documentclass[aps,pra,showpacs]{revtex4}

\begin{document}

\title{Minimum-error discrimination between a pure and a mixed two-qubit state} 

\author{Ulrike Herzog}   

\address{Institut f\"ur Physik,  Humboldt-Universit\"at 
   Berlin, Newtonstrasse 15, D-12489 Berlin, Germany}

\begin{abstract}
The problem of discriminating with minimum error between two mixed quantum 
states is reviewed, with emphasize on the detection operators necessary for 
performing the measurement. 
An analytical result is derived for the minimum probability of
errors in deciding whether the state of a quantum system is either a given
pure state or a uniform statistical mixture of any number of mutually
orthogonal states. The result is applied to two-qubit states,  
and the minimum error probabilities achievable by collective and local measurements 
on the qubits are compared.
\end{abstract}


\maketitle

\section{Introduction}

Quantum state discrimination \cite{chefrev} is of fundamental importance for
quantum communication and 
quantum cryptography. The problem consists in determining the state 
of a single copy of a quantum system that is prepared in a certain but unknown state, 
belonging to a given finite set of known states which occur with given 
a-priori probabilities. 
When the quantum states are non-orthogonal, it is impossible to 
device a measurement that can discriminate between them perfectly. 
Therefore strategies for an optimal measurement have been developed with respect to various 
criteria. Unambiguous discrimination can be achieved at the expense of the 
occurrence of inconclusive results \cite{chefrev} the probability of which is minimized
in the optimum strategy. On the other hand, when a 
conclusive outcome is to be returned in each 
single measurement, 
errors are unavoidable. The strategy for minimum-error discrimination is 
optimized in such a way that
the probability of errors takes its smallest possible value \cite{hel}.
Recently quantum state discrimination has been investigated in the context of 
distinguishing between sets of pure states, or between mixed states, 
respectively. In particular, it has been 
assumed that the actual state of the system belongs to either one of two 
complementary sets of pure states, where each pure state occurs 
with a given a-priori probability. 
Minimum-error discrimination between two sets containing both an arbitrary number 
of pure states has been treated analytically   
under the restriction that the total Hilbert space 
collectively spanned by the states is only two-dimensional \cite{HB}.
If the first set contains only a single state, 
the discrimination problem is referred to as 
quantum state filtering [3 - 5]. 
For optimum unambiguous discrimination, 
general analytical solutions have been derived
to this problem \cite{SBH,BHH}.   
Another recent development consists in studying state discrimination for 
multipartite systems. 
Non-orthogonal bipartite and multipartite states have been considered 
with respect to both minimum-error discrimination and optimum 
unambiguous discrimination [6 - 9]. 
It has been found \cite{virmani} that  
any two pure non-orthogonal multipartite states 
can be discriminated with minimum error using only local 
measurements and classical communication, and that the same holds true for 
two mixed states provided these states span collectively only a 
two-dimensional Hilbert space. 

In the present contribution we consider the problem  
of deciding with minimum error whether the state 
of a quantum system is either a given pure state, 
or a uniform statistical mixture of any number of states
being mutually orthogonal. 
The study is motivated by two main aspects. First, 
it provides another example for an 
analytically solvable minimum-error state discrimination 
problem in an arbitrary dimensional Hilbert space, where so far 
non-trivial explicit solutions have been obtained 
only for discrimination between multiple states that are highly 
symmetric [10 - 13], 
or between three mirror-symmetric
pure states \cite{andersson}.
Second, the solution can be applied to gain some insight into 
the problem of discriminating bipartite quantum states with minimum error. 
Minimum-error discrimination has been 
discussed previously for the joint polarization states of 
two indistinguishable photons travelling in the same spatial mode, 
where the associated Hilbert space is three-dimensional \cite{herzog}. 
Here we shall focus our interest on two-qubit states that span 
a four-dimensional Hilbert space. 
These states could be for instance experimentally realized  
with the help of two polarization-entangled photons  
travelling along different paths. 
We study 
the minimum error probabilities for state discrimination 
that are achievable by collective 
measurement on the two qubits, on the one hand, 
and by local single-qubit measurements, on the other hand. 

\section{Minimum-error discrimination between two mixed states}

We start by briefly reviewing the general problem of 
discriminating with minimum errror between 
two mixed states of a quantum system, 
being characterized by 
the densitity operators $\hat{\rho}_1$ and $\hat{\rho}_2$, and occurring 
with the a-priori probabilities $p_1$ and $p_2$, respectively,  
where $p_1 + p_2 = 1$ \cite{hel,virmani}.   
The corresponding measurement can be formally described  
with the help of two 
positive-semidefinite detection operators,   
$\hat{\Pi}_1$  and $\hat{\Pi}_2$, defined in such a way that 
${\rm Tr}(\hat{\rho}\hat{\Pi}_k)$ is the probability to infer the 
system to be in the state $\hat{\rho}_k$ ($k = 1,2$) if it 
has been prepared in a state $\hat{\rho}$ \cite{chefrev,hel}.
The total probability to get an erroneous result in the measurement 
is given by 
$P_{\rm err}=p_1{\rm Tr}(\hat{\rho}_1\hat{\Pi}_2) + 
p_2{\rm Tr}(\hat{\rho}_2\hat{\Pi}_1)$. 
Since the measurement is exhaustive it follows that  
$\hat{\Pi}_1 + \hat{\Pi}_2 = \hat{1}_{D_S}$, 
with $D_S$ being the dimensionality of the physical state space 
associated with the quantum system 
under consideration. Therefore we obtain 
$P_{\rm err}= p_1 + {\rm Tr}(\hat{\Lambda}\hat{\Pi}_1)
= p_2 - {\rm Tr}(\hat{\Lambda}\hat{\Pi}_2)$,   
where we introduced the Hermitean operator 
\begin{equation}
\hat{\Lambda}= p_2 \hat{\rho}_2 - p_1 \hat{\rho}_1
= \sum_{k=1}^{D_S} \lambda_k |\phi_k\rangle \langle \phi_k|.
\label{1}
\end{equation}
Here the states $|\phi_k\rangle$ denote the 
orthonormal eigenstates belonging to the eigenvalues 
$\lambda_k$. 
By using the spectral decomposition of 
$\hat{\Lambda}$ we get the representations
\begin{equation}
P_{\rm err}=p_1 + 
\sum_{k=1}^{D_S} \lambda_k \langle \phi_k |\hat{\Pi}_1|\phi_k\rangle 
=p_2 - \sum_{k=1}^{D_S} \lambda_k \langle \phi_k |\hat{\Pi}_2|\phi_k\rangle.
\label{2}
\end{equation}
The eigenvalues are real, and without loss of 
generality we can number them in such a way that
$\lambda_k   <  0$ for $k < k_0$, and 
$\lambda_k   >  0$ for $k_0 \leq k \leq D $, where 
$D\leq D_S$, implying that $\lambda_k = 0$ for 
$k > D$. 
The optimization task consists in determining the specific
operators $\hat{\Pi}_1$, or 
$\hat{\Pi}_2$, respectively, that minimize $P_{\rm err}$
under the constraint that 
$0\leq \langle \phi_k |\hat{\Pi}_i|\phi_k\rangle \leq 1$ 
($i=1,2$) for all eigenstates $|\phi_k\rangle$. 
The latter requirement is due to the fact that 
${\rm Tr}(\hat{\rho}\hat{\Pi}_i)$ denotes a probability for any 
$\hat{\rho}$. From this constraint and from 
(\ref{2}) it immediately follows   
that the smallest possible error probability, 
$P_{\rm err}^{\rm min} \equiv P_E$, is achieved 
when 
the equations $\langle \phi_k |\hat{\Pi}_1|\phi_k\rangle = 1$
and $\langle \phi_k |\hat{\Pi}_1|\phi_k\rangle = 0$ 
are fulfilled for eigenstates belonging to negative eigenvalues, 
while eigenstates corresponding to positive eigenvalues
obey the equations 
$\langle \phi_k |\hat{\Pi}_1|\phi_k\rangle = 0$
and $\langle \phi_k |\hat{\Pi}_1|\phi_k\rangle = 1$. 
This is the case when 
\begin{equation}
\hat{\Pi}_{1}   =  \sum_{k=1}^{k_0 -1} 
|\phi_k\rangle \langle \phi_k|,
\qquad
\hat{\Pi}_{2}   =  \sum_{k=k_0}^{D} 
|\phi_k\rangle \langle \phi_k|.
\label{3}
\end{equation}
According to (\ref{2}) the error probability does not change when 
the detection operators are supplemented by projection 
operators onto eigenstates belonging to
the eigenvalue $\lambda_k = 0$, in such a way that 
$\hat{\Pi}_{1} + \hat{\Pi}_{2} = \hat{1}_{D_S}$.
From (\ref{2}) and (\ref{3}) we get
$P_E = p_1 - \sum_{k=1}^{k_0-1}|\lambda_k|
= p_2 - \sum_{k=k_0}^{D}|\lambda_k|$.
By taking the sum of these two alternative representations, using
$p_1 + p_2 = 1$, 
we arrive at the 
well known result \cite{hel}   
\begin{equation}
 P_E = \frac{1}{2}\left(1 - \sum_{k}|\lambda_k|\right) 
   =\frac{1}{2} - \frac{1}{2}\|p_2 \hat{\rho}_2 - p_1 \hat{\rho}_1\|,
\label{4}
\end{equation}
where $\|\hat{\Lambda}\| = 
{\rm Tr}\sqrt{\hat{\Lambda}^{\dag} \hat{\Lambda}}$.
Provided that there are 
positive as well as negative eigenvalues in the 
spectral decomposition of $\hat{\Lambda}$, 
the minimum-error measurement for discriminating 
two quantum states is a von Neumann measurement that 
consists in performing projections onto 
two orthogonal subspaces, as becomes obvious from (\ref{3}). 
On the other hand, when negative eigenvalues do not exist, 
it follows that $\hat{\Pi}_1= 0$
and $\hat{\Pi}_2=\hat{1}_{D_S}$.
Hence the minimum error probability 
can be achieved by always guessing the quantum system to 
be in the state $\hat{\rho}_2$,
without performing any 
measurement at all.   
Similar considerations hold true in the absence of positive 
eigenvalues.  
These findings are in agreement 
with the recently gained insight \cite{hunter} that 
measurement does not always aid minimum-error discrimination.

\section{Distinguishing a pure state and a uniformly mixed state}

Now we apply the general solution, given by (\ref{3}) and (\ref{4}),
to the problem of deciding with minimum error 
whether an arbitrary single-partite or multi-partite 
quantum system is prepared either  
in a given pure state, $|\psi\rangle$,
or in a uniformly mixed state, $\hat{\rho}_2$, 
i. e. we wish to discriminate between the density operators
\begin{equation}
\hat{\rho}_1= |\psi\rangle \langle \psi|,
\qquad 
\hat{\rho}_2= \frac{1}{d}\sum_{j=1}^{d}|u_j\rangle \langle u_j|,
\label{5}
\end{equation}
where $\langle u_i|u_j\rangle = \delta_{ij}$ and 
$d \leq D_S$. 
In the special case $d=D_S$, the state 
$\hat{\rho}_2$ is the maximally mixed state that 
describes a completely random state of the system, containing no 
information at all. Discriminating between $|\psi\rangle\langle\psi|$
and $\hat{\rho}_2$ then amounts to deciding whether the state $|\psi\rangle$
has been reliably prepared, or whether the preparation has totally failed
\cite{hunter}. 
Note that a density operator of the form $\hat{\rho}_2$
would result e. g. if the system was known to be 
prepared with the same a-priori probability, 
$\eta= p_2/d$, in each single one of the states  
$|u_1\rangle\, \ldots, |u_d\rangle$. 
Therefore the solution of our problem 
coincides with the solution of the corresponding quantum state filtering problem.
Without any prior knowledge, however, the detection of  
the state $\hat{\rho}_2$ does not give any information about the method 
used for its preparation. 

In the following we restrict ourselves to the situation that 
$p_1 =  p_2/d$ which means that in the corresponding quantum state 
filtering scenario all possible pure states 
would have equal a-priori probabilities, given by $\eta = 1/(d+1)$.
According to (\ref{4}), the minimum error probability is then determined 
by the eigenvalues $\lambda$ of the operator
\begin{equation}
\hat{\Lambda}= \frac{1}{d+1}\left(\sum_{j=1}^{d}|u_j\rangle \langle u_j|\;-\; 
|\psi\rangle \langle \psi|\right).
\label{6}
\end{equation}
In order to treat the resulting eigenvalue equation,   
\begin{equation}
\hat{F}(\lambda)= \lambda (d+1) \hat{1}_{d+1} +  |\psi\rangle\langle\psi|
-\sum_{j=1}^{d} |u_j\rangle\langle u_j|=  0, 
\label{7}
\end{equation}
we introduce an additional   
basis vector $|u_0\rangle$ in such a way that
\begin{equation}
|\psi\rangle = |u_0\rangle  \sqrt{1-\|\psi^{\parallel}\|^2}\;
+ |\psi^{\parallel}\rangle, 
\qquad
\langle u_0|u_j\rangle = 
\delta_{0,j}.
\label{8}
\end{equation}
Obviously $|\psi^{\parallel}\rangle$  
is the component of 
$|\psi\rangle$ that lies in the subspace spanned by the states  
$|u_1\rangle,\ldots,|u_{d}\rangle$, i. e. 
\begin{equation}
\|\psi^{\parallel}\|^2 = 
\langle \psi^{\parallel}|\psi^{\parallel}\rangle =
\sum_{j=1}^{d}|\langle u_j|\psi\rangle|^2.  
\label{9}
\end{equation}
The total Hilbert space spanned by the set of states 
$\{|\psi\rangle,|u_1\rangle,\ldots,
|u_{d}\rangle\}$ is $d$-dimensional for $\|\psi^{\parallel}\|=1$, and
$d+1$-dimensional for $\|\psi^{\parallel}\|<1$. 
Since $\sum_{j=1}^{d} |u_j\rangle\langle u_j|=
\hat{1}_{d}$ and $\hat{1}_{d+1} = 
\hat{1}_{d} + |u_0\rangle\langle u_0|$,
we find that the eigenvalues $\lambda$ obey 
the equation 
\begin{equation}
{\rm det}(\hat{F}) = {\rm det}(\hat{F_1}) +
{\rm det}(\hat{F_2}) = 0,
\label{10}
\end{equation}
where 
$\hat{F_1}(\lambda) =   |\psi^{\parallel}
\rangle\langle\psi^{\parallel}|
+ [(d+1)\lambda-1]\hat{1}_{d}$ and
$\hat{F_2}(\lambda)  =   |\psi\rangle\langle\psi|
+ [(d+1)\lambda-1]\hat{1}_{d+1}$.
The decomposition (\ref{10}) can be verified by considering the matrix elements 
of $\hat{F}$ in 
the orthonormal basis system $\{|u_0\rangle, \ldots |u_d\rangle \}$ and by expanding   
both ${\rm det}(\hat{F})$ and ${\rm det}(\hat{F_2})$ with respect to their
first rows in this basis. 
We now use the alternative representation
$\hat{1}_{d}= \|\psi^{\parallel}\|^{-2}   
|\psi^{\parallel}\rangle\langle\psi^{\parallel}|
+ \sum_{j=1}^{d-1}|\tilde{u}_j\rangle\langle \tilde{u}_j|$, where 
the $\{|\tilde{u}_j\rangle\}$ are new basis vectors with 
$\langle \psi^{\parallel} |\tilde{u}_j\rangle = 0$, and similiarly, 
we write $\hat{1}_{d+1}=  
|\psi\rangle\langle\psi|+
\sum_{j=1}^{d}|\tilde{v}_j\rangle\langle \tilde{v}_j|$, where 
$\langle \psi |\tilde{v}_j\rangle = 0$. 
Thus we obtain   
\begin{eqnarray}
{\rm det}(\hat{F_1}) & = & \left[ \|\psi^{\parallel}\|^2
+ (d+1) \lambda -1\right]\left[(d+1)\lambda -1\right]^{d-1} \nonumber\\
{\rm det}(\hat{F_2}) & = & (d+1)\lambda\left[(d+1) \lambda -1\right]^{d},
\label{11}
\end{eqnarray}
and upon substituting these expressions into (\ref{10}) we find the eigenvalues
\begin{eqnarray}
\lambda_{1} & = & - \frac{1}{d+1}\sqrt{1-\|\psi^{\parallel}\|^2},\\
\lambda_{2} & = & - \lambda_{1},
\qquad
\lambda_k  =  \frac{1}{d+1}
\qquad
( k=3,\ldots d+1). \nonumber
\label{12}
\end{eqnarray} 
By applying (\ref{4}), the minimum error probability 
follows to be
\begin{equation}
P_E = \frac{1}{d+1}\left( 1 -  \sqrt{1
- \|\psi^{\parallel}\|^2}\right).
\label{13}
\end{equation}
If the density operators to be 
discriminated are linearly independent, i. e. if 
$\|\psi^{\parallel}\| \neq 1$, there
exists exactly one negative eigenvalue, given by $\lambda_1$. 
Therefore the minimum-error measurement is a 
von-Neumann measurement that can be described by 
the detection operators  
$\hat{\Pi}_1 = |\phi_1\rangle\langle\phi_1|$ and 
$\hat{\Pi}_2 =\hat{1}_{D_S} -\hat{\Pi}_1$, where 
$|\phi_1\rangle$ is the eigenstate belonging to $\lambda_1$.
On the other hand, when  $\hat{\rho}_1$ and 
$\hat{\rho}_2$ are linearly dependent, 
i. e. when $\|\psi^{\parallel}\| = 1$,   
a negative
eigenvalue does not exist, and $\hat{\Pi}_2=\hat{1}_{D_S}$. 
In this case the resulting 
minimum error 
probability, $P_E = 1/(d+1)$,
is achievable by guessing the system always to be in the 
state $\hat{\rho}_2$, without 
performing any measurement at all.

It is interesting to
compare the minimum probability of errors, $P_E$, 
with the smallest possible failure probability, $Q_F$, 
that can be obtained  
in a strategy optimized for unambiguously discriminating between 
the quantum states given in (\ref{5}). 
The solution of the latter problem coincides with 
the solution to the problem of optimum unambiguous 
quantum state filtering,   
where the state of the quantum system is known to be either  
$|\psi\rangle$, or any state out of the set of pure states 
$\{|u_1\rangle, \ldots |u_d\rangle\}$. 
The general solution for optimum unambiguous quantum state filtering 
has been provided in \cite{BHH}.   
Assuming equal a-priori-probabilities $\eta = 1/(d+1)$ for all states,
this solution can be directly applied to our case, 
yielding the failure probability $Q_F = 2\,\| \psi^{\parallel}\|/(d+1)$.
By comparing this result with (\ref{13}) it 
becomes obvious that $P_E \leq Q_F$,
where the equality sign holds for $\| \psi^{\parallel}\|= 0$, 
i. e. when $\hat{\rho}_1$ and $\hat{\rho}_2$ are orthogonal.

\section{Application to bipartite qubit states} 

In the following we apply the results of the previous section 
in order to study 
state discrimination for the simplest case of bipartite 
quantum states, i. e. for two-qubit-states. In particular,
we are interested in the question as to what is the difference between 
the smallest possible error probabilities for discriminating 
the two given states, achievable by 
collective measurements on the two qubits, on the one hand, and by a 
local measurement on a single qubit, on the other hand. 
An arbitrary bipartite two-qubit-state, shared among two parties A (Alice) 
and B (Bob), can be expressed with the help of the four   
orthonormal basis states  
\begin{equation}
|v_1\rangle = |00\rangle, \;\; |v_2\rangle= |01\rangle,\;\;
|v_3\rangle = |10\rangle,\;\;|v_4\rangle = |11\rangle,
\label{14}
\end {equation}
where $|mn\rangle$ stands for $|m\rangle_A\otimes|n\rangle_B$, with $|0\rangle$ 
and $|1\rangle$ denoting any two orthonormal basis states of a single qubit. 
In the most general form, the state 
$|\psi\rangle$ and an arbitrary set of four orthonormal states $|u_j\rangle$ 
read
\begin{equation}
|\psi\rangle  =  \sum_{k=1}^4 a_k |v_k\rangle, 
\qquad
|u_j\rangle  =  \sum_{k=1}^4 c_{jk} |v_k\rangle. 
\label{15}  
\end{equation}
Here $j = 0, \ldots,3$, and from normalization, together with the requirement
of orthogonality, it follows that  
\begin{equation}
\sum_{k=1}^4 |a_k|^2 = 1,
\qquad
\sum_{j=0}^3 c_{jk}c_{jl}^{\ast} = \delta_{kl}.
\label{16}    
\end{equation}
Since the state space corresponding to the two-qubit-system 
is four-dimensional, the Ansatz (\ref{5}) is only possible 
when $d \leq 4$. 
Like in the previous section,  
we assume again that $p_1 = 1/(d+1)$.
If $d=4$, the states  
composing $\hat{\rho}_2$ span the entire state 
space of the two-qubit-system and hence 
$\|\psi^{\parallel}\|=1$ for any state $|\psi\rangle$. 
This means that the minimum error probability,  
$P_E = 1/5$, can be achieved by always guessing the 
system to be in the state $\hat{\rho}_2$, 
and there exists no measurement,
neither collective nor local, 
that would lead to a smaller error probability. 
For $d = 3$, the minimum error probability
follows from (\ref{13}) to be 
\begin{equation}
P_E = \frac{1}{4} - \frac{1}{4} \sqrt{1 -
\sum_{j=1}^{3} \left|\sum_{k=1}^4 c_{jk}^{\ast}a_k\right|^2}.
\label{17}
\end{equation}
As an interesting special case we consider the problem 
that Alice and Bob want to decide whether the 
quantum state in question is either the pure state  
$|\psi\rangle \langle \psi|$, or a uniform mixture of the 
three symmetric states 
$|u_{1}\rangle = |00\rangle, 
|u_{2}\rangle = |11\rangle, 
 |u_3\rangle = (|01\rangle + |10\rangle)/\sqrt{2}$.
We then find from (\ref{17}) that 
the minimum error probability is given by 
$P_E=\frac{1}{4}\left(1-\frac{1}{\sqrt{2}}|a_2-a_3|\right)$, 
and the same result would hold true if $|u_1\rangle$ and $|u_2\rangle$
were replaced by the two symmetric Bell states 
$(|00\rangle \pm |11\rangle)/\sqrt{2}$.
According to (\ref{12}) and to the discussion in connection with (\ref{4}), 
minimum-error discrimination is achieved by performing a projection 
measurement onto the eigenstate $|\phi_1\rangle$ that belongs to the  
negative eigenvalue $\lambda_1$ of the operator $\hat{\Lambda}$. 
In general, this eigenstate will be a superposition of 
the two-qubit states (\ref{14}).
The optimum measurement strategy therefore requires a correlated 
measurement that has to be carried out collectively on the two qubits.    

Now we turn to the case that only local measurements are performed, 
and that Alice and Bob are not able to communicate with each 
other. 
Alice wants to distinguish between
the density operators given in (\ref{5}) 
with the smallest possible error that is achievable 
by performing a local measurement on her qubit. 
This means that she has to discriminate with minimum error between
the reduced density operators 
$\hat{\rho}_1^A={\rm Tr}_B (\hat{\rho}_1) $ and 
$\hat{\rho}_2^A={\rm Tr}_B (\hat{\rho}_2) $,
and the minimum error probability takes the form  
\begin{equation}
P_E^{\rm loc} =  
\frac{1}{2} - \frac{1}{2} ||p_2 \hat{\rho}_2^A - 
p_1 \hat{\rho}_1^A|| =
 \frac{1}{2}\left(1 - \sum_{k=1}^2 |\lambda_k^A| \right). 
\label{18}
\end{equation}
Supposing again that $p_1 = 1/(d+1)$, 
the eigenvalues $\lambda_1^A$ and $\lambda_2^A$ refer to 
the operator 
\begin{equation}
\hat{\Lambda}^A= \frac{1}{d+1}\left[\sum_{j=1}^{d} 
{\rm Tr}_B(|u_j\rangle \langle u_j|) 
-{\rm Tr}_B(|\psi\rangle \langle \psi|)\right]
\label{19}
\end{equation}
that acts in the two-dimensional Hilbert space spanned 
by the basis vectors of a single qubit. 
They can be expressed as  
\begin{equation}
\lambda_{1,2}^A =  \frac{L_{00}+L_{11}}{2} \mp 
\sqrt{\frac{(L_{00}-L_{11})^2}{4} + |L_{01}|^2}, 
\label{20}
\end{equation}
where $L_{m_1 m_2} = \langle m_1|\hat{\Lambda}^A|m_2\rangle$
with $\{|m_1\rangle,|m_2\rangle\}= \{|0\rangle_A,|1\rangle_A\}$. 
Using (\ref{14}) -- (\ref{16}) we get the matrix elements 
\begin{eqnarray}
L_{00} & = & \frac{1}{d+1} \sum_{k=1}^2 
        \left(\sum_{j=1}^{d}|c_{jk}|^2  - |a_k|^2 \right),\nonumber\\ 
L_{01} & = & 
    \frac{1}{d+1}\left[\sum_{j=1}^{d}
                  (c_{j1}c^{\ast}_{j3}+c_{j2}c^{\ast}_{j4})  
- (a_1a^{\ast}_3 + a_2a^{\ast}_4)\right],\nonumber\\ 
L_{11} & = & \frac{1}{d+1} \sum_{k=3}^4 
        \left(\sum_{j=1}^{d}|c_{jk}|^2  - |a_k|^2 \right),
\label{21}
\end{eqnarray}
where $d \leq 4$. Obviously, $L_{00} + L_{11} = (d-1)/(d+1)$.
Let us again consider the case the case $d = 3$. In order to 
calculate $P_E^A$ we have to estimate whether 
$\lambda_1^A$ is positive or negative. 
For this purpose we represent the matrix elements 
with the help of the vector $|u_0\rangle$,
making  use of the conditions (16), 
and obtain 
\begin{eqnarray}
L_{00}L_{11}& = & \frac{1}{16}\sum_{k=1}^2 
        \left( |c_{0k}|^2  + |a_k|^2 \right)
        \sum_{k=3}^4 
        \left( |c_{0k}|^2  + |a_k|^2 \right),\nonumber\\
 | L_{01}|^2  
 & = &  \frac{1}{16}|c_{01}c^{\ast}_{03}+c_{02}c^{\ast}_{04}  
+ a_1a^{\ast}_3 + a_2a^{\ast}_4|^2.
\label{22}
\end{eqnarray}
By applying
the Schwarz inequality, 
it can be immediately seen that $|L_{01}|^2 \leq
L_{00}L_{11}$, and it follows that both 
$\lambda_1^A$ and $\lambda_2^A$ cannot be negative. 
Therefore from (\ref{18}) and (\ref{20}) we arrive at
$P_E^{\rm loc} = 1/4$. Taking into account (\ref{13}),
this yields the inequality      
\begin{equation}
P_E^{\rm loc}= \frac{1}{4}\;\;\; \geq \;\;\;P_E
= \frac{1}{4}\left( 1 -  \sqrt{1
- \|\psi^{\parallel}\|^2}\right).
\label{23}
\end{equation}
Obviously, except for the case that $\|\psi^{\parallel}\| = 1$,
i. e. that $\hat{\rho}_1$ and $\hat{\rho}_2$  are linearly dependent, 
the minimum probability of errors achievable 
by a local measurement is larger than the minimum error probability 
resulting from a collective measurement. 
According to the discussion in connection with (4), 
it follows from the positivity of the two eigenvalues that for $d=3$  
and $p_1 = 1/4$ there does not exist any local measurement that gives a 
probability of errors being smaller than the error probability that
would arise from guessing the quantum 
system to be always in the state $\hat{\rho}_2$.
We still mention that for discriminating locally a pure two-qubit state
from a mixture of only two orthogonal two-qubit states, i. e. for $d=2$,  
it is possible that the two eigenvalues have a different sign.
Therefore in this case more subtle investigations are necessary,
in dependence on the specific choice of the
two orthogonal states.          
Moreover, it still remains to be 
studied to what extent the error probability decreases
when classical communication is allowed in addition to  
local measurements.   

\section{Conclusions}

In this paper we investigated the minimum probability of
errors in deciding whether the state of a quantum system is either a given
pure state or a uniform statistical mixture of any number of mutually
orthogonal states. 
Based on our analytical result, we discussed   
the minimum error probabilities achievable by collective and local measurements 
on the two-qubit states.
As a possible application, 
we note that the problem treated in the paper is of particular 
interest 
in the context of quantum state comparison \cite{jex,jex1}, 
where one wants to determine 
whether the states of quantum systems are identical or not. 
It has been shown \cite{jex1} that for comparing two unknown 
single-particle states 
it is crucial to discriminate the anti-symmetric state of the 
combined two-particle system from the 
uniform mixture of the mutually orthogonal symmetric states.   
Finally, it is worth mentioning that 
completely mixed states are important for  
estimating the quality of a source of quantum states, as  
has been recently discussed in connection with single-photon 
sources \cite{hockney}. 
 

\end{document}